\documentclass[cameraready]{Interspeech}

\title{Multi-Level Privacy-Preserving Dementia Detection from Speech via Targeted Adversarial Obfuscation and Representation Learning}

\author[affiliation={1}, orcid=0009-0009-2733-9104, correspondingauthor]
{Henriette Flore}{Kenne}

\author[affiliation={1}, orcid=0000-0001-8216-3353]{Raphael}{Anaadumba}

\author[affiliation={1}]{Mohammad Arif Ul}{Alam}

\address{
    $^1$ Richard A Miner School of Computer and Information Sciences, 
    University of Massachusetts Lowell, Lowell, USA 
}

\email{HenrietteFlore\_Kenne@student.uml.edu, raphael\_anaadumba@student.uml.edu, mohammadariful\_alam@uml.edu}


\usepackage{comment}

\usepackage{amsmath}
\usepackage{algorithm}
\usepackage{algpseudocode}
\usepackage{graphicx}
\usepackage{xcolor}
\usepackage{soul}
\usepackage{booktabs}

\begin{document}

\maketitle

\begin{abstract}
Speech recordings used for dementia detection inherently expose speaker identity, raising critical privacy concerns. Existing methods typically address only singular threats and fail to resolve the privacy--utility trade-off. We propose a multi-level framework designed to neutralize two distinct eavesdropping vectors. At the \textbf{signal level}, a Cumulative Signal Attack (CSA) concentrates perturbations in keyword-aligned regions to maximize transcription error (Word Error Rate $\text{WER} = 1.00$) while preserving vital prosodic biomarkers. At the \textbf{feature level}, a Gradient Reversal Layer (GRL) with Mutual Information (MI)-guided noise injection suppresses speaker-discriminative dimensions while retaining dementia-relevant diagnostic structure. Evaluated on the DementiaBank Pitt Corpus, our framework achieves near-chance speaker identification (Equal Error Rate $\text{EER} = 0.59$, $F1 = 0.003$) while maintaining strong dementia classification performance ($F1 = 0.78$, $\text{AUC} = 0.86$).
\end{abstract}


\section{Introduction}

Dementia, including Alzheimer's disease, is an escalating global health concern, where early detection is essential for timely intervention and improved patient outcomes~\cite{gbd2019dementia2022}. Speech analysis has gained attention as a scalable and non-invasive biomarker of cognitive decline, since neurological impairment manifests in measurable changes in prosodic and acoustic characteristics~\cite{mahon2024voice}. Features such as speech rate, pitch variability, and pause distribution have been shown to reliably distinguish individuals with cognitive impairment from healthy controls~\cite{Ding2024, Pastoriza2022}. However, speech recordings shared for clinical assessment may be intercepted by machine or human eavesdroppers. Voice signals inherently encode both diagnostic biomarkers and personally identifiable information, including vocal tract characteristics and speaking rhythm~\cite{mahon2024voice, bu2025privacy}. This dual-use nature creates a fundamental privacy utility conflict: the same acoustic features that enable dementia detection can also be exploited to re-identify patients or reconstruct transcript content, potentially violating data protection regulations such as Health Insurance Portability and Accountability Act (HIPAA) and the General Data Protection RegulationG (DPR).

Prior work has addressed this through adversarial representation learning~\cite{woszczyk2024prosody}, signal-level perturbation~\cite{Zhong2025}, and embedding-space shuffling~\cite{Arasteh2024}. Woszczyk et al.~\cite{woszczyk2024prosody} proposed a prosody-based adversarial model to obscure speaker embeddings while preserving dementia-relevant features, while others have investigated Gaussian noise injection~\cite{Zhong2025} or feature masking for general-purpose anonymization. However, these approaches suffer from persistent limitations: they either degrade dementia classification performance, leave residual identity cues in high level representations, or lack fine-grained control over the privacy-utility trade-off~\cite{Suhas2023, AlHossain2023}. Critically, most methods operate at a single stage of the processing pipeline, leaving complementary attack surfaces unprotected.

To address these issues, we propose a multi-level privacy-preserving framework operating at both signal and feature levels. At the signal level, we introduce a CSA-based targeted adversarial obfuscation pipeline inspired by~\cite{zhu26} that couples projected gradient descent with CSA to concentrate perturbations within keyword-aligned temporal regions, corrupting ASR transcription while preserving low-frequency prosodic biomarkers. At the feature level, a shared encoder trained with a GRL suppresses speaker-discriminative information in latent representations, complemented by a mutual information-guided selective noise injection mechanism that identifies and protects embedding dimensions most correlated with dementia-relevant prosodic features while systematically corrupting speaker-identifying dimensions. The framework is evaluated on the DementiaBank Pitt Corpus~\cite{Au2023, PMID:8198470} with comprehensive white-box attack assessment. Our main contributions are:
\begin{itemize}
    \item A CSA-based adversarial obfuscation pipeline designed to disrupt semantic content in keyword-aligned regions while retaining prosodic patterns associated with dementia.
    
    \item A mutual information guided selective noise injection strategy that enables dimension-level control of the privacy–utility trade-off in speaker embeddings.
    
    \item Experimental evaluation on the DementiaBank Pitt Corpus assessing privacy preservation and dementia classification performance under multiple attack settings, including white-box adversarial attacks.
\end{itemize}

\vspace{-.1in}
\section{Background}
\subsection{Adversarial Perturbations for Privacy Protection}

Adversarial perturbations, originally introduced as threats to deep learning models~\cite{szegedy2013intriguing, goodfellow2014explaining}, have been repurposed as active privacy mechanisms for speech. By solving 
\begin{align}
    x_{\text{adv}} = x + \epsilon \cdot \text{sign}(\nabla_x \ell(x, y))
\end{align}
where $x$ is the input, $y$ the true label, and $\epsilon$ the perturbation
scale. Imperceptible waveform-level perturbations can be crafted to mislead downstream systems. In the audio domain, Carlini and Wagner~\cite{carlini2018audio} demonstrated targeted ASR manipulation under $\ell_\infty$ constraints, while Qin~\textit{et al.}~\cite{qin2019imperceptible} introduced psychoacoustic masking to ensure perturbations remain inaudible. Building on this, Zhu~\textit{et al.}~\cite{zhu26} proposed CSA-shaped obfuscation to produce temporally smooth perturbations that disrupt ASR transcription while preserving prosodic structure.

\subsection{Threat model}

Voice recordings encode both diagnostic biomarkers and personally identifiable information, creating a dual-use risk that may violate HIPAA and GDPR. We define a dual eavesdropping threat model in which an untrusted third party intercepts clinician-shared data and mounts two attacks: a machine eavesdropper exploits pretrained ASR systems to recover transcript content; a human eavesdropper leverages speaker recognition models to re-link ECAPA-TDNN embeddings or prosodic vectors to individual dentities~\cite{dawalatabad2021ecapa}. We further account for white-box adversaries with partial or full model knowledge~\cite{zhou2024model}. An effective defense must corrupt ASR transcription, suppress speaker-discriminative representations, and preserve clinically meaningful prosodic features~\cite{speaker_adversarial_attacks2024}.
\vspace{-.1in}
\section{Methods}

\begin{figure*}[ht]

  \centering
  \includegraphics[width=\linewidth]{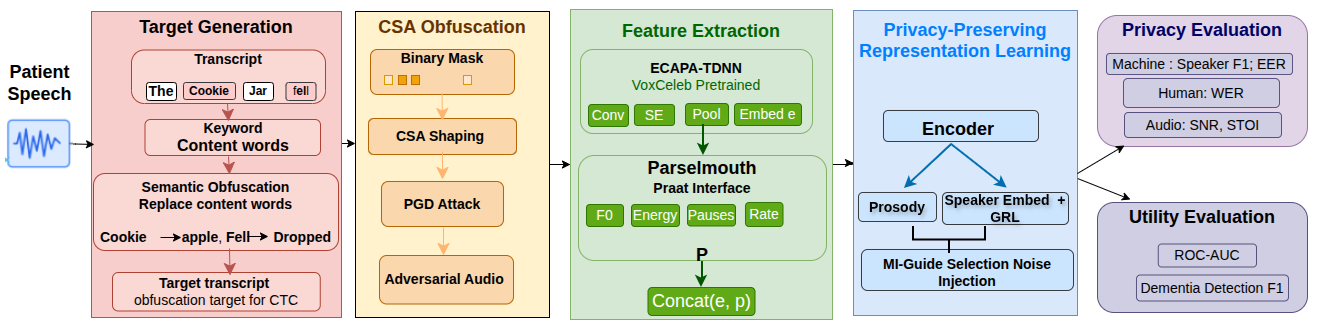}
  \caption{Overview of the proposed multi-level privacy-preserving framework: target generation and CSA-based signal obfuscation (left), feature extraction via ECAPA-TDNN and Parselmouth (centre), and adversarial MI-guided representation learning with privacy and utility evaluation (right)}
  \label{fig:dia}
  \vspace{-0.15in}
\end{figure*}

Our framework addresses the dual eavesdropping threat at two levels, as shown in Fig.~\ref{fig:dia}: (1) a signal-level CSA-based obfuscation pipeline disrupting ASR transcription in keyword-aligned regions while preserving prosodic biomarkers, and (2) a feature-level adversarial representation learning stage suppressing speaker-discriminative dimensions while retaining dementia-relevant structure. While generic perturbations provide coarse obfuscation, they fail to mitigate semantic leakage through ASR a critical vulnerability in clinical speech where transcripts may expose patient identity.

\subsection{CSA-Based Targeted Adversarial Obfuscation}
We design a semantic obfuscation pipeline that couples 
Connectionist Temporal Classification (CTC)-driven adversarial optimisation with a CSA perturbation shaping strategy, repurposing adversarial crafting as a privacy mechanism rather than a threat. Given a clean utterance, we first generate a semantically divergent target transcript by replacing content words (nouns, 
verbs) while preserving dementia-relevant disfluencies such as hesitations, fillers, and pauses. We then minimise the CTC loss with respect to this obfuscated target via projected gradient descent (PGD), iteratively reshaping the waveform to steer the ASR decoder away from the true transcript. Unlike the reversible 
embedding framework of~\cite{10095173}, our approach is strictly non-reversible: the protected waveform is the final output, and no recovery component is employed.

In keyword-aligned perturbation regions, for each utterance, we first identify semantically informative keywords from the transcript. Keyword indices are selected using a content-word filtering strategy and mapped to approximate temporal windows proportional to utterance duration. These windows define the set of patch indices $\mathcal{W}$ where perturbations are permitted. For patch-wise compression, let $\mathbf{x} \in \mathbb{R}^{T}$ denote the clean waveform and $\delta \in \mathbb{R}^{T}$ the raw perturbation updated under an $\ell_\infty$ constraint $\Vert\delta\Vert_\infty \le \epsilon$.  
We partition $\delta$ into non-overlapping patches of size $p$, yielding $K = \lceil T/p \rceil$ patches. The patch-compressed representation is:

\begin{equation}
u_k = \frac{1}{p} \sum_{t=kp}^{(k+1)p-1} \delta_t, 
\quad k = 0, \dots, K-1.
\end{equation}

Only patches aligned to keyword windows ($k \in \mathcal{W}$) are retained; all others are suppressed.

In CSA shaping, Instead of directly applying patch perturbations, we perform cumulative aggregation:
\vspace{-0.15in}
\begin{equation}
c_k = \sum_{i=0}^{k} m_i u_i,
\end{equation}
where $m_k \in \{0,1\}$ is a binary mask indicating keyword-aligned patches. This discrete-time integration promotes low-frequency structure and attenuates abrupt oscillations.

The cumulative signal is expanded back to the waveform domain:
\vspace{-0.15in}
\begin{equation}
\eta_t = c_{\lfloor t/p \rfloor}.
\end{equation}

The CSA-shaped perturbation is clipped to the $\ell_\infty$ bound:

\begin{equation}
\eta = \mathrm{clip}(\eta, -\epsilon, +\epsilon).
\end{equation}

In the adversarial waveform generation, at each PGD iteration, the adversarial waveform is constructed as:

\begin{equation}
\mathbf{x}_{adv} = \mathrm{clip}(\mathbf{x} + \eta, -1, +1),
\end{equation}

We use Wav2Vec2~\cite{10.5555/3495724.3496768} as our differentiable ASR surrogate. Given the clean waveform $\mathbf{x}$, the ground-truth transcript $y$, and the obfuscated target transcript $y_{\mathrm{tgt}}$, we optimise the raw perturbation $\delta$ via projected gradient descent to minimise the CTC loss with respect to $y_{\mathrm{tgt}}$, subject to $\Vert\delta\Vert_\infty \leq \epsilon$:

\begin{equation}
\mathcal{L}_{CTC}(f_{\mathrm{ASR}}(\mathbf{x}_{adv}), y_{\mathrm{tgt}}).
\end{equation}

By acting as a discrete time integration operator within keyword-aligned segments, CSA attenuates high-frequency perturbation components while concentrating adversarial energy in the low-frequency bands most influential for ASR decoding. This reduces perceptual artifacts while maintaining effective ASR misguidance.
The final waveform $\mathbf{x}_{adv}$ is used directly for speaker anonymization evaluation and downstream dementia classification. 


\subsection{Feature Extraction for Utility and Privacy Evaluation}


\textbf{Prosodic Feature Extraction.}
Prosodic characteristics are strongly associated with cognitive decline, encoding patterns of hesitation, speech rhythm, pitch instability, and energy variation that constitute core biomarkers for dementia detection. We extract utterance-level prosodic descriptors from each adversarial waveform using Parselmouth~\cite{JADOUL20181}, yielding a four dimensional feature vector comprising mean fundamental frequency $(F0)$, mean intensity, number of detected pause segments, and articulation rate. Fundamental frequency is estimated via autocorrelation-based pitch tracking; pause segments are identified from unvoiced regions in the pitch contour; and articulation rate is computed as the ratio of voiced frames to total utterance duration. 

\textbf{Speaker Embedding Extraction.}
To quantify speaker identity leakage following CSA obfuscation, we extract speaker embeddings using a pretrained ECAPA-TDNN model~\cite{desplanques2020ecapa} trained on VoxCeleb. ECAPA-TDNN encodes speaker-discriminative characteristics including vocal tract resonances and spectral envelope patterns into a compact 192-dimensional representation. Each adversarial waveform is resampled to 16\,kHz and amplitude-normalised prior to inference. The resulting embedding vectors are used exclusively for privacy evaluation: we assess how well a supervised classifier can recover speaker identity from the obfuscated representations, with a chance-level performance indicating successful anonymisation. 

\subsection{Adversarial Representation Learning}
\label{sec:Adversarial}

We implement an adversarial learning framework \cite{8757975} to suppress speaker-identifiable information while preserving dementia-relevant prosodic characteristics. The architecture consists of a shared encoder with two task-specific branches: a prosody prediction head optimized to reconstruct clinically relevant features, and a speaker classification head trained adversarially via a GRL. Given input feature vector $\mathbf{x} \in \mathbb{R}^d$ combining ECAPA-TDNN embeddings and prosodic features, a shared encoder $f_\theta$ maps $\mathbf{x}$ to a latent representation $\mathbf{z} = f_\theta(\mathbf{x}) \in \mathbb{R}^h$, which feeds two parallel branches: a Prosody Reconstruction: $\hat{\mathbf{y}}_p = h_p(\mathbf{z})$ predicts prosodic features for cognitive assessment.
and a Speaker Classification $\hat{\mathbf{y}}_s = h_s(\text{GRL}(\mathbf{z}, \lambda))$ operates through GRL for identity suppression. The GRL behaves as identity during forward propagation but multiplies gradients by $-\lambda$ during backpropagation, encouraging representations that confuse the speaker classifier while maintaining prosodic information. The system optimizes a composite loss:

\begin{equation}
\mathcal{L}_{\text{total}} = \alpha\,\mathcal{L}_{\text{prosody}} + (1-\alpha)\,\mathcal{L}_{\text{speaker}}^{\text{GRL}},
\end{equation}

where $\mathcal{L}_{\text{prosody}}$ = $\frac{1}{N}\sum_i \Vert h_p(\mathbf{z}_i) - \mathbf{y}_p^i\Vert_2^2$ is the prosody reconstruction Mean Squared Error(MSE) and $\mathcal{L}_{\text{speaker}} = -\frac{1}{N}\sum_i\sum_k y_s^{i,k}\log h_s(\mathbf{z}_i)_k$ is speaker cross-entropy over $K$ classes and $N$ is the number of training utterances.\\

\textbf{Mutual Information-Guided Noise Injection.}
Adversarial training may leave residual speaker information in dimensions where speaker and prosodic characteristics overlap. To address this, we selectively corrupt embedding dimensions according to their prosodic relevance. For each dimension $j$, we compute cumulative mutual information with the prosodic features: $\text{MI}_j = \sum_{k=1}^{K} I(\mathbf{Z}_j;\,\mathbf{P}_k)$. The top 20\% of dimensions by MI score are preserved; the remaining 80\% receive additive Gaussian noise:

\vspace{-.2in}
\begin{equation}
\tilde{Z}_{ij} =
\begin{cases}
Z_{ij}, & j \in \text{ImportantDims}\\
Z_{ij} + \epsilon_{ij}, & \text{otherwise}
\end{cases}, \quad \epsilon_{ij} \sim \mathcal{N}(0,\,\sigma_j^2),
\end{equation}

where $\sigma_j$ is the empirical standard deviation of dimension $j$ scaled by a noise factor of $0.6$. This ensures diagnostically informative dimensions are retained while speaker-discriminative components are systematically degraded.

 \vspace{-0.15in}
\section{Experiments}

\textbf{Dataset.} We evaluate our framework on the DementiaBank Pitt Corpus~\cite{Au2023,PMID:8198470}, comprising 552 Cookie Theft picture description recordings~\cite{Giles01091996} (309 Alzheimer’s disease, 243 healthy controls). Manual transcripts are used exclusively for WER-based privacy evaluation.

\textbf{Implementation.} Experiments were implemented in PyTorch and conducted on an NVIDIA A40 GPU. We used \texttt{facebook/wav2vec2-base-960h} as the differentiable ASR surrogate for CTC-based adversarial optimization. Speaker embeddings were extracted with pretrained ECAPA-TDNN (\texttt{speechbrain/spkrec-ecapa-voxceleb}), and prosodic features were obtained via Parselmouth. Privacy evaluation employed Whisper~\cite{radford2023whisper} for ASR and an RBF-SVM for speaker identification.

\textbf{Metrics.} Audio quality was assessed using SNR, SI-SDR, STOI~\cite{taal2011stoi}, and PESQ~\cite{Rix2001PerceptualEO}. Utility was measured by macro F1-score and ROC-AUC on a stratified 80/20 split. Privacy was evaluated via speaker F1-score and (EER; chance level = 0.5). Semantic obfuscation was quantified using WER between original transcripts and Whisper outputs (WER = 1.0 indicates complete corruption).

\textbf{Hyperparameters.} Parameters were selected via grid search on a validation split. For CSA-based optimization, we explored $\epsilon \in \{0.005, 0.01, 0.02\}$, step size $\alpha \in \{0.005, 0.01, 0.02\}$, and iterations $I \in \{50, 100, 200\}$, optimizing jointly for high WER and preserved STOI. The final configuration ($\epsilon = 0.01$, $\alpha = 0.01$, $I = 100$, $p = 5$) achieved WER = 1.0 with STOI $\geq 0.76$, balancing privacy and intelligibility. For adversarial representation learning, we tuned the prosody speaker trade-off $\alpha \in \{0.3, 0.5, 0.7\}$ and GRL coefficient $\lambda_0 \in \{1, 2, 3\}$, selecting $\alpha = 0.5$ with a scheduled $\lambda$ (2.0 → 1.0). MI-guided noise injection was evaluated with preservation ratios $\{10\%, 20\%, 30\%\}$ and noise scales $\sigma \in \{0.4, 0.6, 0.8\}$; a 20\%/80\% split with $\sigma = 0.6$ provided the best privacy–utility balance.
\vspace{-.1in}

\section{Results}



\subsection{Comparison with Privacy-Preserving Baselines}
Table~\ref{tab:adress-baseline} demonstrates our framework over baseline anonymisation strategies, where an RBF-SVM serves as the speaker identification classifier. While shuffle-based methods achieve speaker suppression (F1\,$\approx$\,0.05), they do so at an unacceptable utility cost, with dementia F1 collapsing to as low as 0.64. Our ADV\,+\,PRIV configuration outperforms all baselines, achieving the best privacy-utility trade-off with dementia F1\,=\,0.79,  dementia F1\,=\,0.79 with only a marginal drop from the non-anonymised original (0.83), AUC\,=\,0.86, speaker F1\,=\,0.22, and EER near chance confirming that dual-level protection is strictly superior to single-stage anonymisation strategies.

\vspace{-.1in}

\begin{table}[ht]
  \centering
  \scriptsize
  \setlength{\tabcolsep}{4pt}
  \caption{Comparison of privacy-preserving methods with state-of-the-art on the DementiaBank Pitt corpus.}
  \label{tab:adress-baseline}
    \begin{tabular}{l|c|c|c}
      \toprule
      \textbf{Source: Pitt Corpus} & \textbf{Dementia F1} $\uparrow$ & \textbf{Speaker F1} $\downarrow$ & \textbf{EER} $\uparrow$ \\
      \midrule
      Original                          & \textbf{0.83} & $0.30 \pm 0.02$ & $0.53 \pm 0.03$ \\
      ADV + PRIV                        & \textbf{0.79} & $0.22 \pm 0.02$ & $0.54 \pm 0.03$ \\
      Shuffle$_{\text{Random}}$         & 0.67          & $0.05 \pm 0.02$ & $0.52 \pm 0.02$ \\
      Shuffle$_{\text{Shap}}$           & 0.74          & $0.05 \pm 0.03$ & $0.54 \pm 0.02$ \\
      Shuffle$_{\text{MI\_AD}}$         & 0.72          & $0.06 \pm 0.01$ & $0.53 \pm 0.03$ \\
      Shuffle$_{\text{MI\_pros}}$       & 0.64          & $0.05 \pm 0.02$ & $0.54 \pm 0.05$ \\
      \midrule
    \end{tabular}
  \vspace{-.2in}
\end{table}
\vspace{-.1in}
\subsection{Adversarial Audio Quality evaluation}

Table~\ref{tab:audio_quality} reports signal-level quality metrics for the CSA-obfuscated waveforms. The negative SNR and SI-SDR values confirm that the perturbation is perceptible at the signal level by design the attack deliberately corrupts the waveform to defeat ASR transcription. Importantly, STOI scores of 0.76 and 0.84 for the dementia and control groups respectively indicate that a degree of speech intelligibility is retained, while PESQ scores of $2.02$ and $3.17$ suggest moderate perceptual quality. 

\begin{table}[ht]
  \centering
  \scriptsize
  \setlength{\tabcolsep}{8pt}
  \caption{Adversarial audio quality on the Pitt Corpus.}
  \label{tab:audio_quality}
  \begin{tabular}{l|c|c|c|c}
    \toprule
    \textbf{Group} & \textbf{SNR} & \textbf{SI-SDR} & \textbf{STOI} & \textbf{PESQ} \\
    \midrule
    Dementia & -11.64 & -11.65 & 0.76 & 2.02 \\
    Control  &  -7.41 &  -7.28 & 0.84 & 3.17 \\
    \bottomrule
  \end{tabular}
  \vspace{-.2in}
\end{table}

\subsection{Privacy Protection}

Table~\ref{tab:eavesdropping} demonstrates strong privacy guarantees against both machine and human eavesdroppers. The machine eavesdropper achieves a \textbf{speaker F1-score of 0.0033} and an \textbf{EER of 0.597} near chance level indicating that ECAPA-TDNN embeddings extracted from adversarial waveforms carry virtually no speaker-discriminative information. For the human eavesdropper scenario, Whisper ASR produces a \textbf{WER of 1.00 } on all obfuscated recordings, confirming complete transcript corruption and full semantic obfuscation.

\begin{table}[ht]
  \centering
  \scriptsize
  \setlength{\tabcolsep}{4pt}
  \caption{Eavesdropping robustness on the Pitt Corpus.}
  \label{tab:eavesdropping}
  \begin{tabular}{l|c|c}
    \toprule
    \textbf{Adversary} & \textbf{Metric} & \textbf{Result} \\
    \midrule
    Machine (ID)     & Speaker F1 $\downarrow$ & 0.0033 \\
    Machine (Verif.) & EER $\uparrow$          & 0.4979 \\
    Human (ASR)      & WER $\uparrow$          & 1.00   \\
    \bottomrule
  \end{tabular}
  \vspace{-.2in}
\end{table}

\subsection{Dementia Detection}

Table~\ref{tab:results} reports dementia classification performance across all evaluated models. MLP achieves the highest \textbf{ROC-AUC of 0.871}, while Logistic Regression yields the best \textbf{F1-score of 0.785}. SVM (RBF) and Random Forest offer competitive performance (AUC 0.844), while tree-based ensemble methods (Gradient Boost, XGBoost, LightGBM) show moderate degradation.

\begin{table}[ht]
  \centering
  \scriptsize
  \setlength{\tabcolsep}{4pt}
  \caption{Dementia detection results with privacy metrics on DementiaBank Pitt Corpus dataset.}
  \label{tab:results}
  \begin{tabular}{l|c|c|c|c}
    \toprule
    \multicolumn{5}{l}{\textit{Dementia detection (Pitt Corpus)}} \\
    \midrule
    \textbf{Models} & \textbf{F1} $\uparrow$ & \textbf{AUC} $\uparrow$ & \textbf{Precision} $\uparrow$ & \textbf{Recall} $\uparrow$ \\
    \midrule
    MLP            & 0.761          & \textbf{0.871} & 0.695 & 0.842 \\
    SVM (RBF)      & 0.772          & 0.843          & 0.680 & \textbf{0.894} \\
    Random Forest  & 0.744          & 0.844          & 0.667 & 0.842 \\
    Gradient Boost & 0.701          & 0.801          & 0.625 & 0.789 \\
    Logistic Reg   & \textbf{0.785} & 0.823          & \textbf{0.702} & 0.868 \\
    XGBoost        & 0.719          & 0.812          & 0.627 & 0.842 \\
    LightGBM       & 0.754          & 0.817          & 0.688 & 0.815 \\
    \bottomrule
  \end{tabular}
  \vspace{-.2in}
\end{table}
\vspace{-.1in}
\subsection{Robustness Against Adversarial Attacks}

Table~\ref{tab:attack_robustness} evaluates the framework's resistance to active adversarial attacks targeting the privacy protection layer. Across all four attack strategies. The system achieves near-perfect privacy preservation \textbf{(PP $\geq$ 0.994) }and diagnostic stability \textbf{(DS $\geq$ 0.994)}, with success rates (SR) at or near zero. These results indicate that the CSA-based obfuscation is robust to white-box and knowledge-guided attack scenarios, providing reliable protection under adversarial conditions.\\

\vspace{-.2in}
\begin{table}[ht]
  \centering
  \scriptsize
  \setlength{\tabcolsep}{4pt}
  \caption{Robustness evaluation against adversarial attacks on the Pitt Corpus.}
  \label{tab:attack_robustness}
  \begin{tabular}{l|c|c|c}
    \toprule
    \textbf{Attack Method} & \textbf{SR} $\downarrow$ & \textbf{PP} $\uparrow$ & \textbf{DS} $\uparrow$ \\
    \midrule
    Gradient Model Inversion  & 0.000 & 1.000 & 1.000 \\
    Parameter Exploitation    & 0.006 & 0.994 & 0.994 \\
    Multi-stage Attack        & 0.000 & 1.000 & 1.000 \\
    Knowledge guided Attack   & 0.000 & 1.000 & 1.000 \\
    \bottomrule
  \end{tabular}
  \vspace{-.1in}
\end{table}

\vspace{-.1in}
\section{Ablation Study}

We probe the privacy-utility trade-off against three signal-level baselines: Gaussian noise ($\sigma \in \{0.0009, \ldots, 0.002\}$), frequency masking ($M=3$ Mel bands), and time stretching ($\tau=2.5$). As shown in Table~\ref{tab:privacy_results}, frequency masking achieves the strongest speaker suppression (F1\,=\,0.00, WER\,=\,0.98) at minimal utility cost (dementia F1\,=\,0.77), while EER remains near chance across all methods. Table~\ref{tab:whitebox} confirms white-box robustness: all methods perfectly resist gradient inversion, with defense scores exceeding 0.986 under multi-stage and knowledge-guided attacks; parameter exploitation remains the most challenging vector (defense\,$\geq$\,0.75).

\begin{table}[ht]
\centering
\vspace{-.1in}
\caption{Privacy-preserving techniques performance on DementiaBank Pitt Corpus.}
\vspace{-.1in}
\label{tab:privacy_results}
\resizebox{\linewidth}{!}{%
\begin{tabular}{l|c|c|c|c}
\midrule
\textbf{Technique} & \textbf{Dementia F1 $\uparrow$} & \textbf{Speaker F1 $\downarrow$} & \textbf{EER $\uparrow$} & \textbf{WER $\uparrow$} \\
\midrule
Gaussian ($\sigma=0.0009$) & 0.78 & $0.08 \pm 0.02$ & $0.57 \pm 0.06$ & 0.74 \\
Gaussian ($\sigma=0.0010$) & 0.76 & $0.07 \pm 0.03$ & $0.56 \pm 0.02$ & 0.86 \\
Gaussian ($\sigma=0.0020$) & 0.75 & $0.04 \pm 0.03$ & $0.56 \pm 0.01$ & 0.92 \\
\midrule
Time Stretching ($\tau$)   & 0.78 & $0.16 \pm 0.04$ & $0.54 \pm 0.05$ & 0.86 \\
Frequency Masking          & 0.77 & $0.00 \pm 0.00$ & $0.52 \pm 0.04$ & 0.98 \\
\bottomrule
\end{tabular}
}
\vspace{-.2in}
\end{table}

\begin{table}[ht]
\centering
\vspace{-.1in}
\caption{White-box attack robustness: success rate $\downarrow$ defense score $\uparrow$ on DementiaBank Pitt Corpus.}
\vspace{-.1in}
\label{tab:whitebox}
\resizebox{\linewidth}{!}{%
\begin{tabular}{l|cc|cc|cc|cc}
\midrule
 & \multicolumn{2}{c|}{\textbf{GI}} & \multicolumn{2}{c|}{\textbf{PE}} & \multicolumn{2}{c|}{\textbf{MS}} & \multicolumn{2}{c}{\textbf{KG}} \\
\textbf{Technique} & Succ.$\downarrow$ & Def.$\uparrow$ & Succ.$\downarrow$ & Def.$\uparrow$ & Succ.$\downarrow$ & Def.$\uparrow$ & Succ.$\downarrow$ & Def.$\uparrow$ \\
\midrule
Gaussian ($\sigma=0.0009$) & 0.000 & 1.000 & 0.120 & 0.880 & 0.006 & 0.994 & 0.005 & 0.995 \\
Gaussian ($\sigma=0.0010$) & 0.000 & 1.000 & 0.136 & 0.864 & 0.014 & 0.986 & 0.012 & 0.988 \\
Gaussian ($\sigma=0.0020$) & 0.000 & 1.000 & 0.122 & 0.878 & 0.006 & 0.994 & 0.005 & 0.995 \\
\midrule
Time Stretching ($\tau$)   & 0.000 & 1.000 & 0.250 & 0.750 & 0.004 & 0.996 & 0.003 & 0.997 \\
Frequency Masking          & 0.000 & 1.000 & 0.160 & 0.984 & 0.008 & 0.992 & 0.007 & 0.993 \\
\bottomrule
\end{tabular}
}
\end{table}

\vspace{-.2in}
\section{Conclusion}
This paper proposes a multi-level privacy-preserving framework 
for speech-based dementia detection that jointly addresses machine and human eavesdropping threats. Using CSA-based adversarial obfuscation at the signal level and MI-guided adversarial representation learning at the feature level, our framework effectively suppresses both transcript and speaker identity leakage while preserving clinically meaningful prosodic biomarkers. Experimental results on the DementiaBank Pitt Corpus demonstrate the effectiveness of combining signal and feature-level protection for achieving a favourable privacy-utility trade-off in real-world dementia screening.

\section{Acknowledgments}

No funding was received for conducting this study.

\section{Use of Generative AI Disclosure}

The authors did not use generative artificial intelligence (AI) tools for the generation of scientific content, analysis, results, or conclusions presented in this work.

\bibliographystyle{IEEEtran}
\bibliography{mybib}

\end{document}